\documentstyle[pre,preprint,aps]{revtex}
\tightenlines

\begin{document}
\draft

\title{
Spike propagation for spatially correlated inputs \\
through noisy multilayer networks 
\footnote{E-print: cond-mat/0308043}
}
\author{
Hideo Hasegawa
\footnote{E-mail:  hasegawa@u-gakugei.ac.jp}
}
\address{
Department of Physics, Tokyo Gakugei University,
Koganei, Tokyo 184-8501, Japan
}
\date{\today}
\maketitle
\begin{abstract}

Spike propagation for spatially correlated inputs
in layered neural networks has been investigated 
with the use of a semi-analytical dynamical mean-field 
approximation (DMA) theory recently proposed by the author
[H. Hasegawa, Phys. Rev. E {\bf 67}, 041903 (2003)].
Each layer of the network is assumed to consist of 
FitzHugh-Nagumo neurons
which are coupled by
feedforward couplings.
Applying single spikes to the network
with input-time jitters whose  
root-mean-square (RMS) value 
and the spatial correlation are $\sigma_I$ and $s_I$,
respectively, we have calculated the RMS value ($\sigma_{Om}$)
and the correlation ($s_{Om}$) of
jitters in output-firing times on each layer $m$.
For all-to-all feedforward couplings,
$s_{Om}$ gradually grows 
to a fairly large value as spikes propagate through the layer,
even for inputs without the correlation.
This shows that for the correlation to be in the range of observed
value of 0.1-0.3, we have to take into account noises and
more realistic feedforward couplings.
Model calculations including local feedforward connections
besides all-to-all feedforward couplings
in multilayers subject to white noises,
have shown that in a long multilayer, $\sigma_{Om}$
and $s_{Om}$ converge to fixed-point values which are
determined by model parameters characterizing the multilayer
architecture.
Results of DMA calculations are in fairly good agreement
with those of direct simulations although the computational time of
the former is much smaller than that of the latter.

\end{abstract}

\noindent
\vspace{0.5cm}
%
\section{INTRODUCTION}


In living brains, information is carried by 
spikes which propagate from one cortical area to another area.
It has been controversial how information
is coded in spikes (for review, see \cite{Rieke96}-\cite{Lestienne01}).
One possibility is that information is coded in the number
of spikes within a short time window ({\it rate code})
\cite{Adrian26}.
Indeed, firing activities of motor and sensory neurons have been
reported to vary in response to applied stimuli.
In an alternative {\it temporal code}, on the contrary,
information is assumed to be carried by precise firing times
of neurons \cite{Softky93}-\cite{Stevens98}.
One of problems in the rate coding is that a fairly long time
of tens of milliseconds are required to read out the rate 
for typical firings rate of 10-100 Hz. 
However, human visual systems, for example, have been reported to
quickly classify patterns within 250 ms despite the fact
that at lest ten synaptic stages are involved from
retina to the temporal brain: transmission times
between the two successive stages are no more than 
10 ms on the average \cite{Thorpe96}.
A possible mechanism to speed up
reading of firing rate may be to collect spikes
of many independent neurons in a population 
({\it population code}) \cite{Knight72}\cite{Petersen01},
where many parallel neurons perform the same task
with the inefficient, high redundancy.
On the other hand,
one of problems in the temporal coding is that spikes are vulnerable 
to noise while the rate coding performs robustly but inefficiently.
These issues on coding have been theoretically studied
in a single neuron ensemble. 
It is not clear whether these conclusions may be applied
to multilayer architectures relevant to cortical processing.
    
Studies with the use of multiunit recordings of frontal cortex
of monkeys have shown that a spatiotemporal pattern of 
highly synchronous firings can propagate through
several tens of synaptic connections \cite{Abeles93}.
A simple model accounting for this phenomenon is
a feedforward synfire chain first proposed by 
Abeles \cite{Abeles93}.
Since the synfire chain model was proposed, many studies 
have been made on its properties 
\cite{Herrmann95}-\cite{Hasegawa03}.
In the rate coding, neurons in each layer are expected to
fire in uncorrelated
manner with other neurons in the same layer.
Neurons in a given layer are assumed to compute
the average firing rate of the neurons in the previous layer
in order to generate the output rate which is related to the input rate.
In a feedforward network, however, 
this firing may propagate to the next layer
in synchronous way \cite{Abeles93}, 
which is detrimental for the rate code. 
Diesmann, Gewaltig and Aertsen \cite{Diesmann99} have shown by
simulations of integrate-and-fire 
(IF) neuron model that a pulse packet can propagate 
through the synfire chain if a packet satisfies the condition which
is specified by the two parameters:
one is the number of spikes in a pulse packet
and the other is the root-mean-square
(RMS) value of firing times in a pulse packet. 
The result of Diesmann et al. \cite{Diesmann99} has been confirmed 
by the method of Fokker-Planck equation \cite{Cateau01}. 

It has been not clear whether feedforward networks support
the rate-code or temporal-code hypothesis.
Shadlen and Newsome \cite{Shadlen94,Shadlen98} have claimed
the feasibility of the rate code,
adopting a model in which excitatory and inhibitory synaptic inputs
are assumed to be balanced.
Because of this balanced input, postsynaptic potentials
fluctuate around the resting potential, which yields
random firings in output neurons.
It has been shown that if each pair of output neurons shares
less than 40 percent of input neurons, only a small degree of
synchrony will be developed, 
which assures an feasibility of the rate code.
The efficiency of the rate code transmission in unbalanced
feedforward networks has been also studied \cite{Rossum02}.
Quite recently, however, it has been pointed out that in {\it long}
feedforward networks, input firing rate cannot be transmitted
reliably because the mean firings rate in a deep layer
is independent of input firing rate \cite{Litvak03}.

Studies on feedforward networks have been so far made 
mostly by direct simulations for networks described by 
the simplest IF model.
It is worthwhile to make a more detailed study on feedforward
multilayers by employing more realistic neuron model 
with an analytical method besides simulations.
In a previous paper \cite{Hasegawa03}
(referred to I hereafter), we have developed
the semi-analytical dynamical mean-field approximation (DMA) theory
as an efficient tool dealing with large-scale
FitzHugh-Nagumo (FN) neuron ensembles subject to noises
\cite{FitzHugh61}\cite{Nagumo62}, by extending the moment method
\cite{Rod96}.
Original $2N$-dimensional
stochastic differential equations (DEs)
for a $N$-unit FN neuron ensemble
are transformed to 
$N(2N+3)$-dimensional deterministic DEs for means, variances 
and covariances of local and global variables.
Recently DMA has been successfully applied to neuron ensembles
described by the realistic Hodgkin-Huxley (HH) model 
\cite{Hasegawa03b}\cite{Note1}.
The FN neuron model adopted in I for 
a feedforward network is obtainable
by a simplification of the HH model \cite{FitzHugh61}\cite{Nagumo62}, 
and it is expected to be more realistic than IF model.
We have investigated in I, the spike propagation through 
the network, taking no account of the spatial correlation.
Experimentally,
correlated firings have been observed in a variety of neurons
\cite{Zohary94}-\cite{Jung00}.
It has been reported that the correlation coefficient between cells 
is about 0.12 in V5 of a rhesus monkey \cite{Zohary94},
0.1-0.3 in human motor units of muscles \cite{Matthews96},
and about 0.3 in cat's lateral geniculate nucleus (LGN) \cite{Alonso96} 
and in retinal ganglion cells of rabbits \cite{DeVries99}.
Theoretical studies on the input correlation
have shown that it may yield
a significant effect on 
the firing rate and the variability of outputs 
\cite{Bernander94}-\cite{Kuhn02}.
Calculations with the use of IF model have shown 
that the firing rate of outputs is increased
with increasing the input correlation for low firing rates of inputs
but is decreased for their high firing rates 
\cite{Bernander94}\cite{Stroeve01}\cite{Kuhn02}.
The variability of output spikes of IF model is an increasing function
of the input correlation, whereas that of HH model is
a decreasing function of the input correlation \cite{Feng01}.

These studies have been made for a single neuron ensemble
\cite{Bernander94}-\cite{Kuhn02}.
We expect that the spatial correlation plays
an important role also in multilayer networks.
Although some theoretical studies have investigated the
cross-correlation of spike rates averaged 
over long times \cite{Litvak03},
there have been no calculations of the firing-time correlation  
in multilayers, as far as the author is concerned.
We have developed, in I, a new method calculating
the instantaneous synchronization ratio in neuron ensembles which is
expressed in terms of variances of local and global variables [Eq. (58)].
As will be shown shortly, the calculated correlation 
in multilayers with all-to-all feedforward couplings
subject to weak noises is developed to a fairly large value 
as spikes propagate,
even for inputs without the correlation.
In order that the correlation remains in the range of the observed
value of 0.1-0.3 mentioned above\cite{Zohary94}-\cite{Jung00},
we have to take into account at least two factors:
one is the more detailed connectivity in feedforward couplings besides 
the all-to-all coupling and the other is noises. 
As for the first issue,
we have assumed, in this study, that
our multilayer network includes, 
besides all-to-all couplings,
local couplings in which each neuron in a given layer
receives an input from one neuron in the preceding layer.
All-to-all and local couplings are superimposed with fractions
of $p$ and $1-p$, respectively, where $p$ denotes a parameter expressing 
a degree of all-to-all component in the total feedforward couplings.
As for the second issue,
several conceivable sources of noises have been reported:
(i) cells in sensory neurons are exposed to noisy outer world,
(ii)  ion channels of the membrane of neurons and synaptic transmission
by a release of synaptic vesicles are essentially stochastic, and
(iii) synaptic inputs include leaked currents from neighboring
neurons.
In this study, we have taken account of
white noises which are independently added to all neurons.
Applying spike inputs 
to the first layer of the network
with the spatial correlation in input-time jitters,
we have investigated the effect of the spatial correlation
in multilayer with all-to-all and local feedforward 
couplings subject to independent noises.

The paper is organized as follows:
In Sec. II, we will discuss an adopted multilayer
with feedforward couplings.
By using DMA, 
the RMS value ($\sigma_{Om}$)
and the correlation ($s_{Om}$) of jitters in output firing times
at layer $m$ are expressed as functions of 
the RMS value ($\sigma_I$) and
the correlation ($s_I$) of jitters in input times. 
In Sec. III,
some model calculations are reported of the 
correlated spike propagation through the multilayer
by using DMA theory and direct simulations.
The final Sec. IV is devoted to discussions and conclusions.

\section{Layered networks consisting of FN neurons}

\subsection{Adopted model}

We have adopted $M$-layer neural networks
in which each layer includes $N$-unit FN neurons.
Dynamics of a single FN neuron $j$ (=1 to $N$)
in a given layer $m$ (=1 to $M$)
is described by nonlinear differential equations (DEs) 
given by 
\begin{eqnarray}
\frac{dx_{mj}(t)}{dt} &=& F[x_{mj}(t)]
- c \:y_{mj}(t) + I_{mj}^{(c1)}(t) + I_{mj}^{(c2)}(t)
+I_{mj}^{(e)}(t) + \xi_{mj}(t),  \\
\frac{dy_{mj}(t)}{dt} &=& b \:x_{mj}(t) - d \:y_{mj}(t)+e,
\end{eqnarray}
with
\begin{eqnarray}
I_{mj}^{(c1)}(t)&=&\left( \frac{w_{1}}{N-1} \right)
\;\sum_{k (\neq j)} G(x_{mk}(t)), \\
I_{mj}^{(c2)}(t)&=& (1-\delta_{m1}) \:w_2
\:[\left( \frac{p}{N} \right)
\;\sum_{k} G(x_{m-1 k}(t)) + (1-p)G(x_{m-1 j}(t)) ], \\
I_j^{(e)}(t) &=& \delta_{m1} \:u \:H(t).
\end{eqnarray}
In Eqs. (1)-(5),
$F[x(t)]=0.5\: x(t)\: [x(t)-0.1]\: [1-x(t)]$, 
$b=0.015$, $c=1.0$, $d=0.003$ and $e=0$ 
\cite{Hasegawa03}\cite{Rod96}, and
$x_{mj}$ and $y_{mj}$ denote the fast (voltage)
and slow (recovery) variables, respectively,
of a given neuron $j$ in the layer $m$; 
$I_{mj}^{(c1)}(t)$ in Eq. (3)
denotes the intra-layer couplings with the strength $w_1$,
the sigmoid function $G(x)$ given by
$G(x)=1/[1+{\rm exp}[-(x-\theta)/\chi]]$,
the threshold $\theta$ and the width $\chi$ \cite{Note6};
the first and second terms of $I_{mj}^{(c2)}(t)$ in Eq. (4) stand for 
all-to-all and local couplings, respectively, 
with the inter-layer feed-forward couplings $w_2$,
$p$ denoting
the degree of common inputs to neuron $j$ in the layer $m$ 
from neurons in the preceding layer $m-1$ \cite{Note7};
$I_j^{(e)}$ in Eq. (5) denotes inputs applied to the
first layer with magnitude of $u$ and an arbitrary
function of $H(t)$ whose explicit form will
be specified below [Eq. (8)];
the last term of Eq. (1), $\xi_{mj}(t)$, expresses the
Gaussian white noise given by
\begin{eqnarray}
<\xi_{mj}(t)>&=&0, \\
<\xi_{mj}(t)\:\xi_{nk}(t')> 
&=& \beta^2 \: \delta_{jk}
\:\delta_{mn}\:\delta(t-t'), 
\end{eqnarray}
where 
$\beta$ denotes 
the magnitudes of noises
and the bracket $< \cdot>$ expresses 
the expectation value.

We will study effects of 
the spatial correlation in single spikes
on the propagation of spike inputs.
We adopt an external input $I_j^{(e)}$ in Eq. (5) 
with the alpha function, $\alpha(t)$:
\begin{eqnarray}
H(t) &=& \alpha(t-t_{Ij})=
[(t-t_{Ij})/\tau_s] \:{\rm exp}[1-(t-t_{Ij})/\tau_s] 
\;\Theta(t-t_{Ij}),
\end{eqnarray}
where $\tau_s$ stands for the synaptic time constant
and $\Theta(t)$ the Heaviside function 
given by $\Theta(t)=1$ for $t \geq 0$
and 0 otherwise.
We assume that jitters in input times $t_{Ij}$ in Eq. (8)
obey the Gaussian distribution with means and variance given by 
\begin{eqnarray}
<t_{Ij}>&=&t_{I}, \\
<\delta t_{Ij} \:\delta t_{Ik}> 
&=& \sigma_I^2 \:[\delta_{jk}+ (1-\delta_{jk})\:s_I], 
\end{eqnarray}
where $\delta t_{Ij}=t_{Ij}-t_{I}$,
$\sigma_I$ and $s_I$ denote RMS value and the spatial correlation,
respectively, of input-time jitters.

When an input spike given by Eqs. (5) and (8) is applied to the first layer, 
it may propagate through the multilayer 
in the propagating regime \cite{Diesmann99}.
The firing time of a given neuron $j$ in the layer $m$
is defined as the time when
the fast variable $x_{mj}(t)$ 
crosses the threshold $\theta$ from below:
\begin{equation}
t_{Omj}= 
\{ t \mid x_{mj}(t) = \theta; \dot{x}_{mj} >0 \}.
\end{equation} 
Means, RMS value and the spatial correlation of jitters in output 
firing times on the layer $m$ are given by
\begin{eqnarray}
t_{Om}&=&<t_{Omj}>, \\
\sigma_{Om}^2&=& 
<\delta t_{Omj}^2>, \\
s_{Om}&=& \frac{1}{N(N-1)} \sum_j \sum_{k (\neq j)}
\frac{<\delta t_{Omj} \;\delta t_{Omk}>}
{\sqrt{<\delta t_{Omj}^2><\delta t_{Omk}^2>}},
\end{eqnarray}
where $\delta t_{Omj}=t_{Omj}-t_{Om}$.
We will calculate $\sigma_{Om}$ and $s_{Om}$ as functions of
$\sigma_I$ and $s_I$ for a set of model parameters
by direct simulations and DMA theory, details of the latter 
being discussed in the following subsection.

\subsection{DMA theory}
\subsubsection{Equations of motions}
As in I \cite{Hasegawa03}, 
we first define the global variables for the layer $m$ by
\begin{eqnarray}
X^m(t)&=&\frac{1}{N}\;\sum_{j} \;x_{mj}(t), \\
Y^m(t)&=&\frac{1}{N}\;\sum_{j} \;y_{mj}(t),
\end{eqnarray}
and their averages by
\begin{eqnarray}
\mu_1^m(t)&=&<X^m(t)>,  \\ 
\mu_2^m(t)&=&<Y^m(t)>.
\end{eqnarray}
Next we define variances and covariances 
between local variables in the layers $n$ and $m$, given by 
\begin{eqnarray}
\gamma^{n,m}_{1,1}(t)
&=& \frac{1}{N}\; \sum_{j} <\delta x_{nj}(t)\:\delta x_{mj}(t)>, \\
\gamma^{n,m}_{2,2}(t)
&=& \frac{1}{N}\; \sum_{j} <\delta y_{nj}(t)\:\delta y_{mj}(t)>, \\
\gamma^{n,m}_{1,2}(t)
&=& \frac{1}{N}\; \sum_{j} <\delta x_{nj}(t)\:\delta y_{mj}(t)>, \\
\gamma^{n,m}_{2,1}(t)
&=& \frac{1}{N}\; \sum_{j} <\delta y_{nj}(t)\:\delta x_{mj}(t)>, 
\end{eqnarray}
and those between global variables in layers $n$ and $m$, given by
\begin{eqnarray}
\rho^{n,m}_{1,1}(t)&=& <\delta X^n(t)\:\delta X^m(t)>, \\
\rho^{n,m}_{2,2}(t)&=& <\delta Y^n(t)\:\delta Y^m(t)>, \\
\rho^{n,m}_{1,2}(t)&=& <\delta X^n(t)\:\delta Y^m(t)>, \\
\rho^{n,m}_{2,1}(t)&=& <\delta Y^n(t)\:\delta X^m(t)>, 
\end{eqnarray}
where $\delta x_{mj}(t)=x_{mj}(t)-\mu_1^m(t)$,
$\delta y_{mj}(t)=y_{mj}(t)-\mu_2^m(t)$,
$\delta X^m(t)=X^m(t)-\mu_1^m(t)$
and $\delta Y^m(t)=Y^m(t)-\mu_2^m(t)$.
It is noted that for $n=m$, we get 
$\gamma_{1,2}^{m,m}=\gamma_{2,1}^{m,m}$
and $\rho_{1,2}^{m,m}=\rho_{2,1}^{m,m}$.

In deriving equations of motions,
we have assumed small $\beta$ and $\sigma_I$, 
and the Gaussian distribution of state variables, as in I.
The interlayer correlation between layers,
which was neglected in I, has been taken into account
within the {\it nearest-layer approximation} (NLA)
in which the correlation beyond adjacent layers
is neglected, as given by
\begin{equation}
\rho_{\kappa, \lambda}^{m-\ell,m}=0.
\hspace{2cm}\mbox{for $\ell > 1
\;\;\;\;(\kappa,\;\lambda=1,\;2)$}
\end{equation}
After some manipulations, 
we get the following DEs for $m$=1 to $M$ (arguement $t$ is neglected):
\begin{eqnarray}
\frac{d \mu_1^{m}}{d t}&=&f^{m}_0 + f^{m}_2 \gamma^{m,m}_{1,1} -c \mu_2^m  
+ w_1 g_o^m + \delta_{m1}\:u\:h_0 + (1-\delta_{m1}) \:w_2\:g_0^{m-1}, \\
\frac{d \mu_2^{m}}{d t}&=& b \mu_1^{m} - d \mu_2^{m} +e,  \\
\frac{d \gamma^{m,m}_{1,1}}{d t}
&=& 2 (a^m \gamma^{m,m}_{1,1}- c \gamma^{m,m}_{1,2})
+ 2w_1 g_1^m \zeta_{1,1}^{m,m}
+\beta^2 + 2 X_{1,1}^m, \\
\frac{d \gamma^{m,m}_{2,2}}{d t}
&=& 2 (b \gamma^{m,m}_{1,2}- d \gamma^{m,m}_{2,2}),  \\
\frac{d \gamma^{m,m}_{1,2}}{d t}&=& b \gamma^{m,m}_{1,1}
+ (a^m-d) \gamma^{m,m}_{1,2} - c \gamma^{m,m}_{2,2} 
+ w_1 g_1^m \zeta_{1,2}^{m,m}
+ X_{1,2}^m,  \\
\frac{d \rho^{m,m}_{1,1}}{d t}&=& 2 (a^m \rho^{m,m}_{1,1} - c \rho^{m,m}_{1,2})
+ 2 w_1 g_1^m \rho_{1,1}^{m.m}
+ \left( \frac{\beta^2}{N} \right) +2 Y_{1,1}^m, \\
\frac{d \rho^{m,m}_{2,2}}{d t}&=& 
2 (b \rho^{m,m}_{1,2}- d \rho^{m,m}_{2,2}),  \\
\frac{d \rho^{m,m}_{1,2}}{d t}&=& b \rho^{m,m}_{1,1}
+  (a^m-d)\rho^{m,m}_{1,2} - c \rho^{m,m}_{2,2}
+ w_1 g_1^m \rho_{1,2}^{m,m}+Y_{1,2}^m,  
\end{eqnarray}
with
\begin{eqnarray}
\zeta_{\kappa,\lambda}^{n,m}
&=&\frac{(\rho_{\kappa,\lambda}^{n,m}-\gamma_{\kappa,\lambda}^{n,m}/N)}
{(1-1/N)}, \\
X_{1,1}^m &=& \delta_{m1}\: u \:h_1 \:P_{1}(t)
+ (1-\delta_{m1})\: w_2 \:g_1^{m-1} 
\:[p\:\rho_{1,1}^{m-1,m}+(1-p) \:\gamma_{1,1}^{m-1,m}], \\
X_{1,2}^m &=& \delta_{m1}\:u \:h_1 \:P_{2}(t)
+ (1-\delta_{m1})\:w_2 \:g_1^{m-1}
\:[p\:\rho_{1,2}^{m-1,m}+(1-p) \:\gamma_{1,2}^{m-1,m}], \\
Y_{1,1}^m &=& \delta_{m1}\:u \:h_1 \:R_{1}(t)
+ (1-\delta_{m1})\:w_2 \:g_1^{m-1} \:\rho_{1,1}^{m-1,m}, \\
Y_{1,2}^m &=& \delta_{m1}\:u \:h_1 \:R_{2}(t)
+ (1-\delta_{m1})\:w_2 \:g_1^{m-1} \:\rho_{1,2}^{m-1,m}, 
\end{eqnarray}
where
$a^m=f_1^m+3 f_3^m \gamma_{1,1}^m$,
$f^m_{\ell} = (1/\ell !) F^{(\ell)}(\mu_1^m)$,
$g^m_{\ell} = (1/\ell !) G^{(\ell)}(\mu_1^m)$ and 
$h_{\ell} = (1/\ell !) d^{\ell} \:H(t)/dt^{\ell}$.
In Eqs. (37)-(40), $P_{\kappa}$ and $R_{\kappa}$ 
($\kappa=1, 2$)
express contributions to the first layer ($m=1$),
obeying the following DEs
(see the Appendix A):
\begin{eqnarray}
\frac{d P_{1}}{dt}&=& a^1 P_{1} - c P_{2} 
+\left( \frac{w_1 g_1^1}{N-1} \right) [NR_{1}-P_{1}] 
+\sigma_I^2 \:u \:h_1, \\
\frac{d P_{2}}{dt}&=& b P_{1} - d P_{2}, \\
\frac{d R_{1}}{dt}&=& a^1 R_{1} - c R_{2} 
+w_1 \:g_1^1\:R_{1} 
+[\frac{1}{N}+(1-\frac{1}{N})\:s_I] \:\sigma_I^2 \:
u \:h_1, \\
\frac{d R_{2}}{dt}&=& b R_{1} - d R_{2}. 
\end{eqnarray}
In Eqs. (37)-(40), 
$\gamma_{\kappa, \nu}^{m-1,m}$ and $\rho_{\kappa, \nu}^{m-1,m}$ 
express the interlayer correlation between
layers $m-1$ and $m$,
satisfying DEs given by
\begin{eqnarray}
\frac{d \gamma_{1,1}^{m-1,m}}{dt} &=&
(a^{m-1}+a^{m})\gamma_{1,1}^{m-1,m}
-c(\gamma_{1,2}^{m-1,m}+\gamma_{2,1}^{m-1,m}) 
+ w_1(g_1^{m-1}+g_1^m) \zeta_{1,1}^{m-1,m} \nonumber \\
&+&w_2 g_1^{m-1}
[p\:\rho_{1,1}^{m-1,m-1}+(1-p)\:\gamma_{1,1}^{m-1,m-1}]
, \\
\frac{\gamma_{2,2}^{m-1,m}}{dt} &=&
b(\gamma_{1,2}^{m-1,m}+\gamma_{2,1}^{m-1,m})-2d \gamma_{2,2}^{m-1,m}, \\
\frac{\gamma_{1,2}^{m-1,m}}{dt} &=&
b \gamma_{1,1}^{m-1,m}+(a^{m-1}-d) \gamma_{1,2}^{m-1,m}
-c \gamma_{2,2}^{m-1,m} 
+ w_1 g_1^{m-1} \zeta_{1,2}^{m-1,m}, \\
\frac{\gamma_{2,1}^{m-1,m}}{dt} &=&
b \gamma_{1,1}^{m-1,m}+(a^{m}-d) \gamma_{2,1}^{m-1,m}
-c \gamma_{2,2}^{m-1,m} 
+ w_1 g_1^m \zeta_{2,1}^{m-1,m}
\nonumber \\
&+&w_2 g_1^{m-1} [p\:\rho_{2,1}^{m-1,m-1}+(1-p)\:\gamma_{2,1}^{m-1,m-1}],\\
\frac{d \rho_{1,1}^{m-1,m}}{dt} &=&
(a^{m-1}+a^{m})\rho_{1,1}^{m-1,m}-c(\rho_{1,2}^{m-1,m}+\rho_{2,1}^{m-1,m})
+w_1(g_1^{m-1}+g_1^{m} )\rho_{1,1}^{m-1,m}\nonumber \\
&+&w_2 g_1^{m-1}\rho_{1,1}^{m-1,m-1}, \\
\frac{\rho_{2,2}^{m-1,m}}{dt} &=&
b(\rho_{1,2}^{m-1,m}+\rho_{2,1}^{m-1,m})-2d \rho_{2,2}^{m-1,m}, \\
\frac{\rho_{1,2}^{m-1,m}}{dt} &=&
b \rho_{1,1}^{m-1,m}+(a^{m-1}-d) \rho_{1,2}^{m-1,m}
-c \rho_{2,2}^{m-1,m} + w_1 g_1^{m-1} \rho_{1,2}^{m-1,m}, \\
\frac{\rho_{2,1}^{m-1,m}}{dt} &=&
b \rho_{1,1}^{m-1,m}+(a^{m}-d) \rho_{2,1}^{m-1,m}
-c \rho_{2,2}^{m-1,m} 
+ w_1 g_1^{m} \rho_{2,1}^{m-1,m}
+w_2 g_1^{m-1} \rho_{2,1}^{m-1,m-1}.
\end{eqnarray}
Original $2 N M$-dimensional deterministic DEs 
given by Eqs. (1)-(5) are transformed to 
$N_{eq}$-dimensional deterministic DEs
given by Eqs. (28)-(52)
where $N_{eq}=12+16 (M-1)$.
If there are no jitters ($P_{\kappa}=R_{\kappa}=0$) 
or if the interlayer correlation
is neglected ($\rho_{\kappa, \lambda}^{n,m}=0$ for $m \neq n$),
DMA leads to $8 M$-dimensional DEs as in I.
We note that
the contribution of input jitters to the first layer
is proportional to $\sigma_I^2$ in Eq. (41) whereas that
to $[1/N+(1-1/N) s_I] \sigma_I^2$ in Eq. (43).

\subsubsection{Quantities relevant to output firings}
\noindent
{\bf Neuron Activity and Firing-Time Distribution}

We will show, in this subsection, that from
$\mu_1^m(t)$, $\gamma_{1,1}^{m,m}(t)$ and $\rho_{1,1}^{m,m}(t)$
which are obtained from Eqs. (28)-(52),
we may calculate the three important quantities
relevant to output firings on the layer $m$:
the activity of neurons ($a_{Om}$), 
the RMS value of output jitters ($\sigma_{Om}$)
and their spatial correlation ($s_{Om}$). 

The averaged distribution of 
the voltage variable $x_{mj}(t)$ 
is described by the Gaussian distribution with the mean
of $\mu_1^m(t)$ and the variance of $\gamma_{1,1}^{m,m}(t)$ 
\cite{Hasegawa03}\cite{Rod96}.
The probability $W_{Om}(t)$ when $x_{mj}(t)$ at $t$
is above the threshold $\theta$ is given by
\begin{equation}
W_{Om}(t)=1 - \psi \left( \frac{\theta-\mu_1^m(t)}
{\sqrt{\gamma_{1,1}^{m,m}(t)}} \right),
\end{equation} 
where $\psi(y)$ is the error function given by an integration
from $-\infty$ to $y$ of the normal distribution
function $\phi(x)$:
\begin{equation}
\phi(x)=\frac{1}{\sqrt{2 \pi}} 
{\rm exp}\left( -\frac{x^2}{2} \right).
\end{equation} 
The neuron activity in the layer $m$ is given by
\begin{equation}
a_{Om} = W_{Om}(t_{Om}^*),
\end{equation}
which is unity when all neurons in the layer fire at
$t=t_{Om}^*$ defined by $\mu_i^m(t_{Om}^*)=\theta$.
The fraction of firings of neurons
in the layer $m$ is given by \cite{Hasegawa03}
\begin{equation}
Z_{Om}(t) = \frac{d W_{Om}}{d t} 
\sim \phi(\frac{t-t^{*}_{Om}}{\sigma_{Om}})\;
\frac{d}{dt}(\frac{\mu_1^m}{\sqrt{\gamma_{1,1}^{m,m}}}) 
\; \Theta(\dot{\mu_1}^m),
\end{equation} 
with the RMS value of jitters of output spikes given by
\begin{equation}
\sigma_{O m}= \sqrt{<\delta t_{Oj}^2>}
=\frac{\sqrt{\gamma_{1,1}^{m,m}}}{\dot{\mu_1}^m},
\end{equation} 
where $\mu_1^m$, $\dot{\mu_1}^m$ and $\gamma_{1,1}^{m,m}$ are evaluated 
at $t_{Om}^*$.
Our $\sigma_{Om}$ corresponds to $\sigma_O$, RMS of firings times,
of Diesmann, Gewaltig and Aertsen \cite{Diesmann99}.

\vspace{0.5cm}
\noindent
{\bf Synchronization ratio and Correlation of output firings}

The synchronization ratio $S_{m}(t)$ in a given layer $m$ 
is given by \cite{Hasegawa03}
\begin{eqnarray}
S_m(t)&=&
\frac{[\rho^{m,m}_{1,1}(t)/\gamma^{m,m}_{1,1}(t)-1/N]}{(1-1/N)} \\
&=&\frac{1}{N(N-1)} \sum_{j}\sum_{k(\neq j)}
\frac{<\delta x_{mj} \:\delta x_{mk}>}
{\sqrt{<\delta x_{mj}^2><\delta x_{mk}^2>}},
\end{eqnarray}
which is 0 and 1 for completely asynchronous and synchronous states,
respectively. 
Then the spatially-averaged correlation of output firing times 
in layer $m$ defined by Eq. (14), is given by
\begin{eqnarray}
s_{Om}&=& \frac{1}{N(N-1)} \sum_j \sum_{k (\neq j)} 
\frac{<\delta t_{Omj}\:\delta t_{Omk}>}
{\sqrt{<\delta t_{Omj}^2><\delta t_{Omk}^2>}}=S_{m}(t_{Om}^*),
\end{eqnarray}
where the relation given by Eq. (57) is adopted.

Thus $a_{Om}$, $\sigma_{Om}$ and $s_{Om}$ given by Eqs. (55), (57) and (60), respectively, are expressed in terms of 
$\mu_{1}^m$, $\gamma_{1,1}^{m,m}$ and $\rho_{1,1}^{m,m}$, and
they depend on model parameters
of $\sigma_I$, $s_I$, $p$, $\beta$, $w_1$, $w_2$ and $N$.

\section{Model calculations}
\subsection{Effects of $s_I$}

In this study, we pay our attention to
the response of multilayer networks
to a single spike input of $I^{(e)}(t)$
with $t_{I}=100$ in Eqs. (5) and (8).
We have adopted the parameters of $u=0.10$,
$\theta=0.5$, $\chi=0.1$ and $\tau_s=5$. 
Parameter values of $\sigma_I$, $s_I$, $p$, $\beta$, $w_1$, $w_2$, 
$N$ and $M$ will be explained shortly.
The value of $u=0.10$ has been chosen
for a study of the response to a supra-threshold input,
because the critical magnitude of $u$ is $u_c=0.0435$ below which
firings of neuron  defined by Eq. (11) cannot take place 
for $\sigma_I=\beta=0$. 
Direct simulations have been performed by solving
$2 M N$ DEs given by  Eqs. (1)-(5) 
with the use of the fourth-order Runge-Kutta method with
a time step of 0.01 for hundred trials
otherwise noticed. 
Correlated input times of $t_{Ij}$ given by Eqs. (9) and (10)
have been generated by the Gaussian-distribution programs.
DEs of DMA given by Eqs. (28)-(52) have been solved by using 
also the fourth-order Runge-Kutta method with
a time step of 0.01.
All calculated quantities are dimensionless. 

Raster in Fig. 1(a) shows firings of neurons 
of the first ten layers in a multilayer
of $N=10$ and $M=20$ for a typical set of parameters of
$\sigma_I=1$, $s_I=0$, $p=1$, $\beta=0.01$, $w_1=0$ and
$w_2=0.1$, calculated by a direct simulation (a single trial).
The ordinate expresses the neuron index $k$ defined
by $k=10(m-1)+j$ where $m=1-10$ and $j=1-10$.
The uppermost cluster denotes firings of ten neurons in the layer $m=1$
and the bottom cluster those in the layer $m=10$.
When spikes are applied at $t=100$, 
neurons are already randomized 
because noises have been applied since $t=0$.
Firings occur with a delay of about 5 at each stage, and
it takes about 48 for spikes to propagate from $m=1$ to $m=10$.
When the noise intensity is increased to $\beta=0.02$,
fluctuations of firings due to inputs and
spurious firings are increased, as shown in Fig. 1(b).
Figures 2(a) and 2(b) show
time courses of $\mu_1^m(t)$ and $S_m(t)$
of the first ten layers for the same set of parameters 
as in Fig. 1(a):
solid curves denote the results of DMA
theory and dashed curves those of direct simulations.
Figure 2(a) shows that a spike propagates from $m=1$ to $m=10$.
Results of $\mu_1^m(t)$ of DMA are in good agreement with
those of direct simulations; the former is not distinguishable
from the latter.
The synchronization ratio of $S_{m}(t)$ shown in Fig. 2(b) is 
zero at $m=1$
because of the vanishing input correlation $s_I=0$.
Nevertheless, $S_{m}(t)$ after receiving inputs
gradually become large as a spike
propagates through the layer.
This development in the synchrony is more clearly realized
in Fig. 3(a), where large open and filled circles, respectively,
show the $m$ dependence of the correlation
of $s_{Om}=S_m(t_{Om}^*)$ for $s_I=0$ and $p=1$ calculated by
direct simulation (dashed curve) and DMA (solid curve).
We note that although $s_{Om}=0$ at $m=1$ for $s_I=0$, 
it is rapidly increased and saturates with 
a value of about 0.71 (0.61) 
in direct simulations (DMA calculation) at $m=20$. 
On the contrary, open and filled squares, respectively,
in Fig. 3(a) show that $s_{Om}$ for $s_I=1$
and $p=1$ is decreased at $m \geq 1$ and again show the saturation 
with a value of about 0.87 (0.71)
in direct simulation (DMA calculation) at $m=20$.
For $0< s_I < 1$, $s_{Om}$ show a similar, gradual change as $m$ 
is increased.
It is noted that the agreement between
the results of DMA calculations and direct simulation is 
good at small $m$ but become worse at larger $m$.
This is due to the adopted NLA in which 
the correlation beyond the nearest layers is neglected.  

It is noted that an increase in the synchrony 
as $m$ is increased,
which is realized for $s_I=0$ and 0.2 in Fig. 3(a),
is due to common inputs arising from
all-to-all interlayer couplings for $p=1$ in Eq. (4). 
In fact, if we set $p=0$ for which inputs come only
through local couplings in Eq. (4), the synchrony is 
gradually decreased as spikes propagate
by effects of random noises for all values of $s_I$, 
as shown in Fig. 3(c).  
In the intermediate $p$ value, for example, for $p=0.4$,
the synchrony is decreased
(increased) compared with that for $p=1$ ($p=0$),
as shown in Fig. 3(b).

Figure 4(a), 4(b) and 4(c) express the $m$ dependence of $\sigma_{Om}$,
RMS value of jitters in firing times,
for $p=1.0$, 0.4 and 0, respectively,
with various $s_I$ values.
Although $s_{Om}$ is variable
depending on $s_I$ and $p$ as shown in Figs. 3(a)-3(c),
magnitudes of $\sigma_{Om}$ are nearly independent of $m$,
which shows that
spikes propagate with nearly the same dispersion. 
In particular, for $p=0$, the $m$ dependence of $\sigma_{Om}$ 
is almost the same for all $s_I$ values, as shown in Fig. 4(c).
Because of the adopted NLA, the agreement between
the results of DMA calculations and direct simulation
become worse at larger $m$ although both results
are similar in the qualitative sense.
The neuron activity $a_{Om}$ defined by Eq. (55) is 0.50 - 0.51
at $m \geq 1$ for all the cases investigated (not shown).

So far we have adopted values of $N=10$ and $M=20$.
It is desirable to perform numerical calculations
with larger values of $N$ and $M$ for a better understanding of
multi-layer networks in living brains. 
Because of a limitation of our computer facility,
we have performed only DMA calculations 
for larger value of $N$ and $M$. 
Figure 5(a), 5(b) and 5(c) show the $m$-dependence of $s_{Om}$
for $p=1.0$, 0.4 and 0.0, respectively,
with $N=100$ and $M=40$ for various $s_I$ values:
parameters of $\beta$, $w_1$, $w_2$ are same as in Fig. 2.
When comparing Figs. 5(a)-5(c) with 2(a)-2(c), we note similar $m$
dependence in them: 
the $N$ dependence of $s_{Om}$ will be shortly discussed 
in Sec. IIID.

\subsection{Effects of $p$}

As was pointed out in Figs. 3(a)-3(c), the factor of $p$ plays an 
important role for synchrony in spike propagation.
In order to systematically study the effect of $p$ on the input-output
relation of $M=20$ multilayer,
we have calculated $s_{O \:20}$, $s_{Om}$ at $m=M=20$,
as a function of $p$
for various $s_I$ values
with $\sigma_I=1$, $\beta=0.01$ and $N=100$,
whose result is shown in Fig. 6.
For $p=0$, $s_{O \:20}$ is very small for all $s_I$.
When $p$ is increased from 0, $s_{O \:20}$ is linearly increased 
and it shows an almost saturation at $p > 0.5-0.6$.

Figure 7(a) depicts the calculated result showing 
$s_{O \:20}$ against $s_I$, the input-output relation of the
correlation for a fixed value of $\beta=0.01$ 
in the multilayer of $M=20$ and $N=100$.  
It is shown that for independent local couplings only ($p=0$),
$s_{O\:20}$ becomes too small compared to $s_I$.
In contrast, for common all-to-all feedforward couplings only
($p=1$), $s_{O\:20}$ 
becomes larger than the input correlation
for $s_I < s_{Ic}$ where $s_{Ic}=0.54$ is the critical value
below which $s_{O\:20} > s_I$. 
For $p=0.2$ and 0.4, the critical value becomes $s_{Ic}=0.09$
and 0.33, respectively,
which are nearly the same as
the experimentally observed value of 0.1-0.3 
\cite{Zohary94}-\cite{Jung00}.

\subsection{Effects of $\beta$}

As was shown in Fig. 1(b), noises are detrimental 
for the synchrony of spikes.
This fact is realized when we compare 
$s_{Om}$ for $\beta=0.02$ shown in Fig. 5(d)
with that for $\beta=0.01$ in Fig. 5(a).
The value of $s_{Om}$ for $s_I=1$ at $m=40$ is about 0.20
in Fig. 5(b) which is much smaller than 0.46 in Fig. 5(a). 

Figure 7(b) expresses $s_I$ versus $s_{O\:20}$ 
when the noise intensity is changed with a fixed value of $p=1$
for $\sigma_I=1$ and $N=100$.
In the case of $\beta=0.01$, $s_{O \:20}$ is larger than $s_I$
for $s_I < s_{Ic}=0.54$. 
On the contrary, in the case of $\beta=0.02$, the critical value 
is $s_{Ic} = 0.18$.
Furthermore, in the case of $\beta=0.03$, we get
$s_{Ic} = 0.11$.
Thus $s_{Ic}$ is much reduced with increasing $\beta$.

\subsection{Effects of $N$}

As mentioned above, $s_{O\:20}$ becomes
smaller than $s_I$ for $s_I > 0.18$ for $\beta=0.02$ and $N=100$.
This situation is changed if the size of $N$ is reduced.
Figure 8(a) shows $s_{Om}$ for various $N$ with $s_I=0.4$ and $\beta=0.02$.
In the case of $N=10$, for example, $s_{O\:20}$ is 0.50 which is larger
than 0.19 for $N=100$. For $N=20$, $s_{O\:20}$ is nearly the
same as $s_I=0.4$. Figure 8(a) clearly shows that 
$s_{O\:20}$ is gradually decreased as $N$ is increased.

\subsection{Effects of $w_1$}

So far we have assumed vanishing intralayer couplings, $w_1$,
which are now introduced.
When intralayer coupling
$w_1$ are positive (excitatory),
$s_{Om}$ is expected to be increased.
This is confirmed in our calculations shown in Fig. 8(b)
depicting $s_{Om}$ for $w_1$=0, 0.05 and 0.1
with $\sigma_I=1$, $s_I=0.4$, $\beta=0.02$, $w_2=0.1$ and $N=100$.
On the contrary, if $w_1$ is negative (inhibitory),
it is considered to prevent the propagation of a spike.
Actually, Fig. 8(b) shows that for $w=-0.05$, a propagation of a spike 
is terminated at $m=7$ below which $s_{Om}$ is smaller 
than that for positive $w_1$.

\subsection{Effects of $w_2$}

The interlayer coupling $w_2$ is expected to 
play also important roles in spike propagation.
Figure 8(c) shows calculated results
when the interlayer coupling $w_2$ is increased from 0.1 to 0.2,
which yields an increase in $s_{Om}$ for
$\sigma_I=1$, $s_I=0.4$, $w_1=0$,
$\beta==0.02$ and $N=100$.
On the contrary, our calculation in Fig. 8(c) shows
that the negative couplings with $w_2=-0.1$ and -0.2
are not favorable for the spike propagation which is
terminated at $m = 9$.

\section{Conclusions and Discussions}

We have discussed the spatial correlation while spikes propagate
through feedforward multilayer.
Figures 3(a) and 3(b) suggest that as $m$ is increased,
$s_{Om}$ and $\sigma_{Om}$ may approach fixed values.
In order to show this more clearly, we depict, in Fig. 9(a), 
the $\sigma_{Om}$-$s_{Om}$ plot in which points of ($\sigma_{Om}$, $s_{Om}$)
are sequentially connected from $m=0$ to $m=40$
with $\beta=0.01$ and $N=100$: note that
($\sigma_{Om}$, $s_{Om}$) for $m=0$ stand for input values.
For example, in the case of $\sigma_I=1$ and $s_I=1.0$, the point starts
from ($\sigma_{Om}$, $s_{Om}$)=(1.0, 1.0) at $m=0$ 
and ends with (0.58, 0.45) at $m=40$.
In contrast, in the case of $\sigma_I=1$ and $s_I=0.0$, the point starts
from (1.0, 0.0) at $m=0$ and ends with (0.49, 0.22) at $m=40$.
In the case of $\sigma_I=0$ and $s_I=0.0$, the point varies
from (0.0, 0.0) at $m=0$ to (0.48, 0.21) at $m=40$.
These show that a fixed point may be about 
$(\sigma_{O\infty}, s_{O\infty}) \sim (0.54, 0.38)$,
as shown by the cross in Fig. 9(a).

Figure 9(b) show a similar $\sigma_{Om}$-$s_{Om}$ plot
for a larger $\beta=0.02$. 
In the case of $\sigma_I=1$ and $s_I=1.0$, for example, the point starts
from (1.0, 1.0) at $m=0$ and ends with (0.95, 0.22) at $m=40$.
In the case of $\sigma_I=1$ and $s_I=0.0$, the point changes
from (10.0, 1.0) at $m=0$ to (0.92, 0.16) at $m=40$.
Results for $\sigma_I=0$ and $s_I=0$
and for $\sigma_I=1$ (and 2) with $0 \leq s_I \leq 1$
show that $(\sigma_{O\infty}, s_{O\infty}) \sim (0.93, 0.18)$
for $\beta=0.02$.

Including calculated results of the neuron activity 
$a_{Om}$ [Eq. (55)], which becomes
$a_{Om}=0.50-0.51$ at $m \geq 1$ for all the cases 
investigated (not shown), 
we may say that
all curves starting from different initial values 
of $\sigma_I$ and $s_I$ converge to the
fixed point of $(a_{O \infty}, \:\sigma_{O\infty}, \:s_{O\infty})$
in the three-dimensional space spanned by $a_{Om}$, $\sigma_{Om}$ and 
$s_{Om}$. 
The fixed point is determined by the parameters characterizing the
multilayer architecture such as $\beta$, $p$, $w_1$, $w_2$ and $N$,
but independently of the parameters of $\sigma_I$ and $s_I$ 
for input signals. 
Our conclusion supplements the result of 
Diesmann {\it et al.}\cite{Diesmann99}
who have shown that in the propagating regime, the number of 
firing neurons and
RMS of firing times in a pulse packet converge to fixed-point values.
Our calculation has shown that 
while a pulse packet propagates with an almost 
constant dispersion (RMS),
the spatial correlation within the packet may change, 
and in a deep layer, it saturates at the value determined by the
parameters depending on the multilayer.

The dependence of calculated 
fixed-point values of $s_{O\infty}$ and $\sigma_{O\infty}$
on $\beta$, $p$, $N$, $w_1$ and $w_2$ are summarized as follows.

\noindent
(1) $\sigma_{O\infty}$ is increased as increasing $\beta$,
but decreased as increasing $p$, $N$, $w_1$ or $w_2$.

\noindent 
(2) $s_{O\infty}$ is increased as increasing $p$,
$w_1$ or $w_2$, but decreased as increasing $\beta$ or $N$.

\noindent
We have tried to elucidate this property by
an analysis using DMA.
Because the fixed points do not depend on $\sigma_I$ and
$s_I$, we consider the case of $\sigma_I=s_I=0$.
In the case of $w_1=w_2=0$, Eqs. (28)-(52) yield
\begin{eqnarray}
\gamma_{1,1}^{m,m} &\propto& \beta^2, \\
\rho_{1,1}^{m,m} &\propto& \frac{\beta}{N},
\end{eqnarray}
for $m \rightarrow \infty$ where they are independent of $m$.
When $w_1$ and $w_2$ are small, Eqs. (30), (33), (37), (39), (45) and (48)
yield following equations given as series of $w_1$ and $w_2$:
\begin{eqnarray}
\gamma_{1,1}&\equiv& {\rm lim}_{m \rightarrow \infty} 
\;\gamma_{1,1}^{m,m} = c\: \beta^2(1-a_1w_1-a_2w_2)
+d \:w_2[p\:\rho_{1,1}^{'}+(1-p)\:\gamma_{1,1}^{'}], \\
\rho_{1,1} &\equiv&  {\rm lim}_{m \rightarrow \infty} 
\;\rho_{1,1}^{m,m} = 
\left( \frac{c\:\beta^2}{N} \right)  (1-b_1w_1-b_2w_2)
+d \:w_2 \rho_{1,1}^{'}, \\
\gamma_{1,1}^{'} &\equiv& {\rm lim}_{m \rightarrow \infty} 
\;\gamma_{1,1}^{m-1,m}
= e \:w_2 \:[p \:\rho_{1,1}+(1-p)\:\gamma_{1,1}], \\
\rho_{1,1}^{'} &\equiv& {\rm lim}_{m \rightarrow \infty}
\;\rho_{1,1}^{m-1,m}
= e \:w_2 \:\rho_{1,1},
\end{eqnarray}
where expansion coefficients of $a_1$, $a_2$, $b_1$, $b_2$
$c$, $d$ and $e$ are obtainable
from Eqs. (28)-(52) in principle although their explicit
forms are not necessary for our qualitative discussion.
Solving Eqs. (63)-(66) for $\gamma_{1,1}$ and $\rho_{1,1}$,
which are substituted to Eqs. (57) and (60), we get
\begin{eqnarray}
\sigma_{O \infty} &\propto&
\left( \frac{\beta}{\dot{\mu_1}} \right)
\left( 1 -\frac{1}{2}(a_1 w_1 +a_2 w_2)
+\frac{1}{2}d\:e\:w_2^2\:[(1-p)^2 + \frac{1}{N}p(2-p)] \right), \\
s_{O \infty} &\simeq &
\left( \frac{a_1-b_1}{N-1} \right) w_1
+ \left( \frac{a_2-b_2}{N-1} \right) w_2
+ \left( \frac{d\:e\:p(2-p)}{N} \right) w_2^2, 
\end{eqnarray}
where only relevant terms are retained.
Expressions given by Eqs. (67) and (68) 
may account for all the dependence of 
$\sigma_{O\infty}$ and $s_{O\infty}$
raised in items (1) and (2), except 
the $N$ dependence of $\sigma_{O\infty}$ and
the $\beta$ dependence of $s_{O\infty}$.
The former may be explained 
if $a_1$ or $a_2$ is an increasing function of $N$, and the latter
if $a_1$ or $a_2$ ($b_1$ or $b_2$) is a decreasing 
(increasing) function of $\beta$.

Although numerical calculations reported in Sec. III have been 
made for single spike inputs, we may easily apply our DMA theory
to the case of spike-train inputs given by [see Eq. (8)]
\begin{equation}
I_{j}^{(e)}(t)= \delta_{m1}\:u \:\sum_{\ell}
\alpha(t-t_{Ij\ell})
\end{equation}
where $t_{Ij\ell}$ stands for the $\ell$th input time to neuron $j$.
Raster in Fig. 10(a) shows firings of neurons
for an applied Poisson spike train with 
the average interspike interval (ISI) of 100
($\beta=0.02$ $\sigma_I=1$, $s_I=0$, $p=1$,  
$w_1=0$, $w_2=0.1$, $M=20$ and $N=10$) 
calculated by direct simulation (a single run).
In contrast, raster in Fig. 10(b) expresses {\it global} firings 
on layer $m$ for which the firing time is defined by
\begin{equation}
t_{Omg}= 
\{ t \mid \mu_1^{m}(t) = \theta; \dot{\mu}_1^{m} >0 \}.
\end{equation} 
where $\mu_1^m(t)$ denotes the averaged voltage
variable on the layer $m$ [Eq. (17)].
Time courses of $\mu_1^m(t)$ are plotted in Fig. 10(c),
in which the uppermost frame shows an applied
Poisson spike train, $I^{(e)}(t)$.
The result of $\mu_1^m(t)$ of DMA (solid curves) is in good agreement
with that of direct simulations (dashed curves).
Figures 10(a)-(c) show that spikes propagate from $m=1$
to $m=10$ of the $M=20$ multilayer.
Spurious firings due to added noises, which is realized in Fig. 10(a),
vanish in global firings shown in Fig. 10(b)
by the averaging over neuron ensembles,
which expresses the {\it population effect} 
\cite{Knight72}\cite{Petersen01}.
Comparing $\mu_1^m(t)$ with an applied spike train 
of $I^{(e)}(t)$, for example, at $t \sim 300-600$, 
we note that when ISI of
input spikes is shorter than about 55, FN neuron cannot respond
because of its refractory period, which is realized also 
in HH neuron \cite{Hasegawa00}.
It is noted that although the correlation for
spike-train inputs develops  
while spikes propagate through the multilayer,
as for single spike inputs,
means and variances of their ISI remain
almost the constant. 

To summarize,
we have studied the spatial correlation during
spike's propagation through multilayers
to show

\noindent
(a) the input correlation of $s_I$ may propagate
through the network, yielding
$s_{Om} \sim s_I$ at the end layer of $m=M \sim 10-20$
with the observed magnitude of $s_{OM} \sim 0.1 - 0.3$ 
\cite{Zohary94}-\cite{Jung00},
when model parameters are appropriate, and

\noindent
(b) in a long multilayer, the correlation of the deep layer
converges to a fixed-point value of ($\sigma_{O\infty},s_{O\infty}$) 
which depends on the parameters characterizing the multilayer architecture
but is independent of
the input correlation.

\noindent
The item (1) implies that spikes in multilayers with physiologically
reasonable size of $m=M \sim 10-20$ may carry information
encoded in the spatial correlation of firing times.
The item (b) is similar to the result obtained in Ref.\cite{Litvak03},
where the spike rate in a deep layer of a balanced synfire chain
is shown to be independent of the input rate.

Finally we would like to point out the efficiency of our DMA.
Direct simulations with 100 trials for a multilayer of
$M=20$ and $N=10$ with a set of parameters (Figs. 1 and 2),
required the computation time of 51 minutes
by using 1.8 GHz CPU PC,  
while a DMA calculation needs only 6 s,
which is about 500 times faster than simulations.    

\section*{Acknowledgements}
This work is partly supported by
a Grant-in-Aid for Scientific Research from the Japanese 
Ministry of Education, Culture, Sports, Science and Technology.

\newpage

\appendix

\section{Derivation of Eqs. (41)-(44)}


From Eqs. (1)-(5), we get DEs for the
deviations of $\delta x_{mj}$ and $\delta y_{mj}$ 
of a neuron $j$ (=1 to $N$)
in the layer $m$ (=1 to $M$),
given by (see Appendix A in I)
\begin{eqnarray}
\frac{d \delta x_{mj}}{d t}&=& f_1^m \delta x_{mj}
+f_2^m (\delta x_{mj}^2-\gamma_{1,1})
+ f_3^m \delta x_{mj}^3 - c \delta y_{mj}  
+ \delta I_{mj}^{(c1)} + \delta I_{mj}^{(c2)} + \delta I_{mj}^{(e)} 
+\xi_{mj}, \\
\frac{d \delta y_{mj}}{d t}&=& b \delta x_{mj} - d \delta y_{mj},
\hspace{1cm}\mbox{($j \in m$)}
\end{eqnarray}
with
\begin{eqnarray}
\delta I_{mj}^{(c1)}
&=& \left( \frac{w_1 g_1^m }{N-1} \right)
\sum_{k(\neq j)} \delta x_{mk}, \\
\delta I_{mj}^{(c2)}
&=& (1-\delta_{m1})\;w_2 g_1^m [\frac{p}{N}
\sum_{k} \delta x_{m-1 k}+(1-p) \delta x_{m-1 j}], \\
\delta I_{mj}^{(e)}
&=& - \delta_{m1} \;u\: \dot{H}\: \delta t_{Ij},
\end{eqnarray}
where $\dot{H}=d H(t)/dt$.
We have taken into account up to third-order terms in $\delta x_{mj}$
which play an important role in stabilizing Des \cite{Hasegawa03},
while only a linear-order term is included in the coupling term in
Eqs. (A3) and (A4).
DEs for the variances and
covariances are given by
\begin{eqnarray}
\frac{d \gamma_{\kappa,\lambda}^m}{d t}
&=& \frac{1}{N} \sum_{j}[
2 < \delta x_{mj} \:(\frac{d \delta x_{mj}}{d t})> 
\; \delta_{\kappa 1} \delta_{\lambda 1}
+ <\delta y_{mj} \:(\frac{d \delta x_{mj}}{d t})
+\delta x_{mj} \:(\frac{d \delta y_{mj}}{d t}) >
\delta_{\kappa 1} \delta_{\lambda 2} \nonumber \\
&+& 2 < \delta y_{mj}\:(\frac{d \delta y_{mj}}{d t}) > 
\; \delta_{\kappa 2} \delta_{\lambda 2} 
], \\
\frac{d \rho_{\kappa,\lambda}^{n,m}}{d t}
&=& \frac{1}{N^2} \sum_{j} \sum_{k} [
2 < \delta x_{nj} \:(\frac{d \delta x_{mk}}{d t}) >
\; \delta_{\kappa 1} \delta_{\lambda 1} 
+ <\delta y_{mk} \:(\frac{d \delta x_{nj}}{d t})
+\delta x_{nj} \:(\frac{d \delta y_{mk}}{d t}) > 
\delta_{\kappa 1} \delta_{\lambda 2} \nonumber \\
&+& <\delta y_{nj} \:(\frac{d \delta x_{mk}}{d t})
+\delta x_{mk} \:(\frac{d \delta y_{nj}}{d t}) > 
\delta_{\kappa 2} \delta_{\lambda 1} 
+ 2 < \delta y_{nj}\:(\frac{d \delta y_{mk}}{d t}) > 
\; \delta_{\kappa 2} \delta_{\lambda 2} 
], 
\end{eqnarray}
Substituting Eqs. (A1)-(A5) to Eqs. (A6) and (A7), we get DEs for 
$\gamma_{\kappa,\lambda}^{n,m}$ and 
$\rho_{\kappa,\lambda}^{n,m}$
($\kappa, \lambda=1,2$).
In the process of these calculations,
we get new correlation functions of $P_{\kappa}(t)$ and
$R_{\kappa}(t)$ defined by
\begin{eqnarray}
P_{\kappa}(t)&=& - \frac{1}{N} \sum_j
(<\delta x_{1j}(t) \:\delta t_{Ij}> \delta_{\kappa 1}
+<\delta y_{1j}(t) \:\delta t_{Ij}> \delta_{\kappa 2}), \\
R_{\kappa}(t)&=& - \frac{1}{N^2} \sum_j \sum_k 
(<\delta x_{1j}(t) \:\delta t_{Ik}> \delta_{\kappa 1}
+<\delta y_{1j}(t) \:\delta t_{Ik}> \delta_{\kappa 2}), 
\end{eqnarray}
whose equations of motions are given by
Eqs. (41)-(44).



\begin{figure}
\caption{
Raster showing firings of neurons on 
the first ten layers in a multilayer of $N=10$ and $M=20$
for (a) $\beta=0.01$ and (b) $\beta=0.02$
with $\sigma_I=1$, $s_I=0$, $p=1$,  
$w_1=0$, $w_2=0.1$, $M=20$ and $N=10$,
calculated by a direct simulation (a single trial).
The vertical scale expresses the neuron index $k$ defined by
$k=10 (m-1)+j$ where $1 \leq m \leq 10$ and $1 \leq j \leq 10$. 
}
\label{fig1}
\end{figure}

\begin{figure}
\caption{
Time courses of (a) $\mu_1^m$ and (b) $S_{om}$
in $1 \leq m \leq 10$ of a multilayer with $N=10$ and 
$M=20$ for $\sigma_I=1$, $s_I=0$, $p=1$, $\beta=0.01$,  
$w_1=0$ and $w_2=0.1$,
calculated by DMA (solid curves) and
direct simulations of 100 trials (dashed curves).
}
\label{fig2}
\end{figure}

\begin{figure}
\caption{
$s_{Om}$ for (a) $p=1.0$, (b) $p=0.4$ and (c) $p=0.0$,
with $\sigma_I=1$, $\beta=0.01$,
$w_1=0$, $w_2=0.1$, $N=10$ and various $s_I$:
$s_I=1.0$ (squares), 0.8 (circles), 0.6 (diamonds),
0.4 (inverted triangles), 0.2 (triangle) 
and 0 (large circles),
calculated by direct simulations (open marks) 
and DMA (filled marks):
}
\label{fig3}
\end{figure}

\begin{figure}
\caption{
$\sigma_{Om}$ for (a) $p=1.0$, (b) $p=0.4$ and (c) $p=0.0$,
with $\sigma_I=1$, $\beta=0.01$,
$w_1=0$, $w_2=0.1$, $N=10$ and various $s_I$:
$s_I=1.0$ (squares), 0.8 (circles), 0.6 (diamonds),
0.4 (inverted triangles), 0.2 (triangle) 
and 0 (large circles),
calculated by direct simulations (open marks) 
and DMA (filled marks):
}
\label{fig4}
\end{figure}

\begin{figure}
\caption{
$s_{Om}$ for 
(a) $\beta=0.01$ and $p=1.0$, 
(b) $\beta=0.01$ and $p=0.4$, 
(c) $\beta=0.01$ and $p=0.0$, and 
(d) $\beta=0.02$ and $p=1$,
with $\sigma_I=1$,
$w_1=0$, $w_2=0.1$, $N=100$ and various $s_I$
calculated by DMA:
$s_I=1.0$ (squares), 0.8 (circles), 0.6 (diamonds),
0.4 (inverted triangles), 0.2 (triangle) 
and 0 (large circles).
}
\label{fig5}
\end{figure}

\begin{figure}
\caption{
The $p$ dependence of $s_{O\:20}$ 
for various $s_I$
with $\sigma_I=1$, $\beta=0.01$, 
$w_1=0$, $w_2=0.1$ and $N=100$
calculated by DMA:
$s_I=0.80$ (circles), 0.6 (diamonds), 0.4 (inverted triangles), 
0.2 (triangle) and 0 (large circles).
}
\label{fig6}
\end{figure}

\begin{figure}
\caption{
(a) $s_{O\:20}$ against $s_I$ for $\beta=0.01$ with various $p$:
$p=1$ (circles), 0.6 (squares), 0.4 (inverted triangles),
0.2 (triangles) and 0 (diamonds).
(b) $s_{O\:20}$ against $s_I$ for $p=1$ with various $\beta$:
$\beta=0.01$ (circles), 0.02 (triangles) and
0.03 (squares).
(a) and (b) are calculated by DMA 
with $\sigma_I=1$, $w_1=0$, $w_2=0.1$ and $N=100$.
}
\label{fig7}
\end{figure}

\begin{figure}
\caption{
(a) $s_{Om}$ for different $N$,
(b) for different $w_1$, and (c) for different $w_2$,
with $\sigma_I=1$, $s_I=0.4$, $p=0.4$, $\beta=0.02$,
calculated by DMA.
}
\label{fig8}
\end{figure}

\begin{figure}
\caption{
$\sigma_{O m}$ against $s_{Om}$ 
for (a) $\beta=0.01$ and (b) $\beta=0.02$,
with $\sigma_I=1$, $p=1$
$w_1=0$, $w_2=0.1$, $M=40$ and $N=100$
calculated by DMA.
Arrows denote the direction of increasing $m$.
All points starting from ($\sigma_{O\:0}, \;s_{O\:0}$)
converge to the fixed-point marked by the cross (see text).
}
\label{fig9}
\end{figure} 

\begin{figure}
\caption{
(a) Raster of firings of individual neurons,
(b) raster of global firings averaged on each layer
in a multilayer 
calculated by a direct simulation (a single trial),
and (c) time courses of $\mu_1^m$ calculated
by DMA (solid curves) and simulations (dashed curves)
for applied Poisson spike inputs $I^{(e)}$ shown at the uppermost frame
of (c):
$\beta=0.02$ $\sigma_I=1$, $s_I=0$, $p=1$,  
$w_1=0$, $w_2=0.1$, $M=20$ and $N=10$.
The vertical scale of (a) expresses the neuron index $k$ defined by
$k=10 (m-1)+j$ where $1 \leq m \leq 10$ and $1 \leq j \leq 10$ (see text). 
}
\label{fig10}
\end{figure}

\end{document}